\documentclass{IAU}
\usepackage{graphicx}

\title[Episodic accretion in focus: revealing the environment of FU\,Orionis-type stars] 
{Episodic accretion in focus: revealing the environment of FU\,Orionis-type stars}

\author[O.\,Feh\'er et al.]   
{O.\,Feh\'er$^{1,2}$,
 \'A.\,K\'osp\'al$^{1,3}$, P. \'Abrah\'am$^1$, M.\,R.\,Hogerheijde$^4$, Ch.\,Brinch$^5$, D.\,Semenov$^3$}

\affiliation{$^1$ Konkoly Observatory, Research Centre for Astronomy and Earth Sciences, \\ Hungarian Academy of Sciences, 1121 Budapest, Konkoly Thege Miklós út, 15-17, Hungary \\[\affilskip]
$^2$ Institut de Radioastronomie Millimétrique, 38400, \\ Saint-Martin-d'Hères,
300 Rue de la Piscine, Grenoble, France \\[\affilskip]
$^3$ Max Planck Institute for Astronomy, Königstuhl 17, D-69117 Heidelberg, Germany \\[\affilskip]
$^4$ Leiden Observatory, Leiden University, Niels Bohrweg 2, 2333 CA Leiden, The Netherlands \\[\affilskip]
$^5$ Niels Bohr International Academy, The Niels Bohr Institute, \\ University of Copenhagen, Blegdamsvej 17, 2100 Copenhagen Ø, Denmark
}

\pubyear{2019}
\volume{345}  
\jname{Origins: from the Protosun to the First Steps of Life}
\editors{Bruce G. Elmegreen, L. Viktor T\'oth, Manuel G\"udel, eds.}
\begin{document}

\maketitle

\begin{abstract}
The earliest phases of star formation are characterised by intense mass accretion from the circumstellar disk to the central star. One group of young stellar objects, the FU\,Orionis-type stars exhibit accretion rate peaks accompanied by bright eruptions. The occurance of these outbursts might solve the luminosity problem of protostars, play a key role in accumulating the final star mass, and have a significant effect on the parameters of the envelope and the disk. In the framework of the Structured Accretion Disks ERC project, we are conducting a systematic investigation of these sources with millimeter interferometry to examine whether they represent normal young stars in exceptional times or they are unusual objects. Our results show that FU\,Orionis-type stars can be similar to both Class\,I and Class\,II systems and may be in a special evolutionary phase between the two classes with their infall-driven episodic eruptions being the main driving force of the transition.
\end{abstract}

\firstsection 
\section{Introduction}

Star formation is a complicated interplay of fragmentation and gravitational collapse of interstellar clouds, the mass accretion onto the surface of the newly formed star, turbulence, magnetic fields and environmental effects. These mechanisms play out on multiple scales and compete with each other, thus to understand them, a self-consistent description is needed. One of the important results in the last few years was that protostars prove to be less luminous than theorized \cite[(Dunham et al. 2014 and references therein)]{dunham2014}. Studies of eruptive young stars e.g. EX\,Lupi (EXors) and FU\,Orionis-type sources (FUors) \cite[(Herbig 1977; Hartmann \& Kenyon, 1996)]{herbig1977} revealed that these Sun-like stars acquire their mass through outburst events. These powerful eruptions can have significant consequences for the luminosity and the final mass of the stars and influence their outflows and the disk structures, see e. g. \cite[Kun et al. (2011), Mosoni et al. (2013), Cieza et al. (2016)]{mosoni2013, kun2011, cieza2016}. 

FUors are low-mass, pre-main sequence stars, showing 5$-$6\,mag brightness increase in the optical and infrared regime, then slow fading or a brightness plateau over years or decades. The eruptions are caused by enhancement of the accretion rate from the circumstellar disk to the star, from a typical rate of 10$^{-7}$ to 10$^{-4}$\,M$_{\odot}$\,yr$^{-1}$. The eruptions were theorised to be triggered by gravitational instability \cite[(Armitage et al. 2001; Boley et al. 2006)]{armitage2001, boley2006}, viscous-thermal instability \cite[(Bell\,\&\,Lin 1994)]{bell1994}, or the effect of a close companion \cite[(Bonnell\,\&\,Bastien 1992)]{bonnell1992}. Recent studies suggest that the outbursts are recurring and that infall from an envelope is needed to replenish the material in the disk \cite[(Vorobyov\,\&\,Basu 2006; Vorobyov et al. 2013)]{vorobyov2006, vorobyov2013}. After a sequence of eruptions the envelope clears out and the star may enter a quiescent phase, first showing smaller magnitude eruptions like EXors, then becoming similar to T\,Tauri-type stars \cite[(Vorobyov\,\&\,Basu 2015)]{vorobyov2015}.

Only a few tens of FUors are known so far \cite[(Audard et al. 2014)]{audard2014}. Analysis of their infrared emission and 10\,$\mu$m silicate feature led to an evolutionary sequence proposed by \cite[Green et al. (2006)]{green2006} and \cite[Quanz et al. (2007)]{quanz2007}. Category\,1 FUors are younger, embedded in dense, dusty envelopes with the silicate feature in absorption, while Category\,2 FUors have a remnant envelope, with the silicate feature in emission. FUor outbursts thus seem to happen during the first few million years of star formation and affect the inner disk heavily. Since this is the site and time of terrestrial planet formation, investigation of the process is imperative to describe the stages of star and planet formation.

\section{Eruptive young stars through millimeter interferometry}

High angular resolution and high sensitivity interferometry offered by the Atacama Large Millimeter/submillimeter Array (ALMA) or the Northern Extended Millimeter Array (NOEMA) allows us to study the morphology and kinematics of FUor disks and envelopes. 
We measured the emission of $^{13}$CO, C$^{18}$O and the 2.7\,mm continuum around 12 FUors with NOEMA, reaching spatial resolutions of around 1000\,au: V1057\,Cyg, V1515\,Cyg, V2492\,Cyg and V2493\,Cyg in 2012, V733\,Cep, V1735\,Cyg and RNO\,1B/C in 2014, and V899\,Mon, V900\,Mon, V960\,Mon, and V582\,Aur in 2017. We also performed observations with the 30\,m telescope to detect the CO emission correctly on larger scales and made a spectral survey in the 110\,GHz band to search for other molecular species.

\begin{figure}[t]
\centering
\begin{center}
 \includegraphics[width=.49\linewidth]{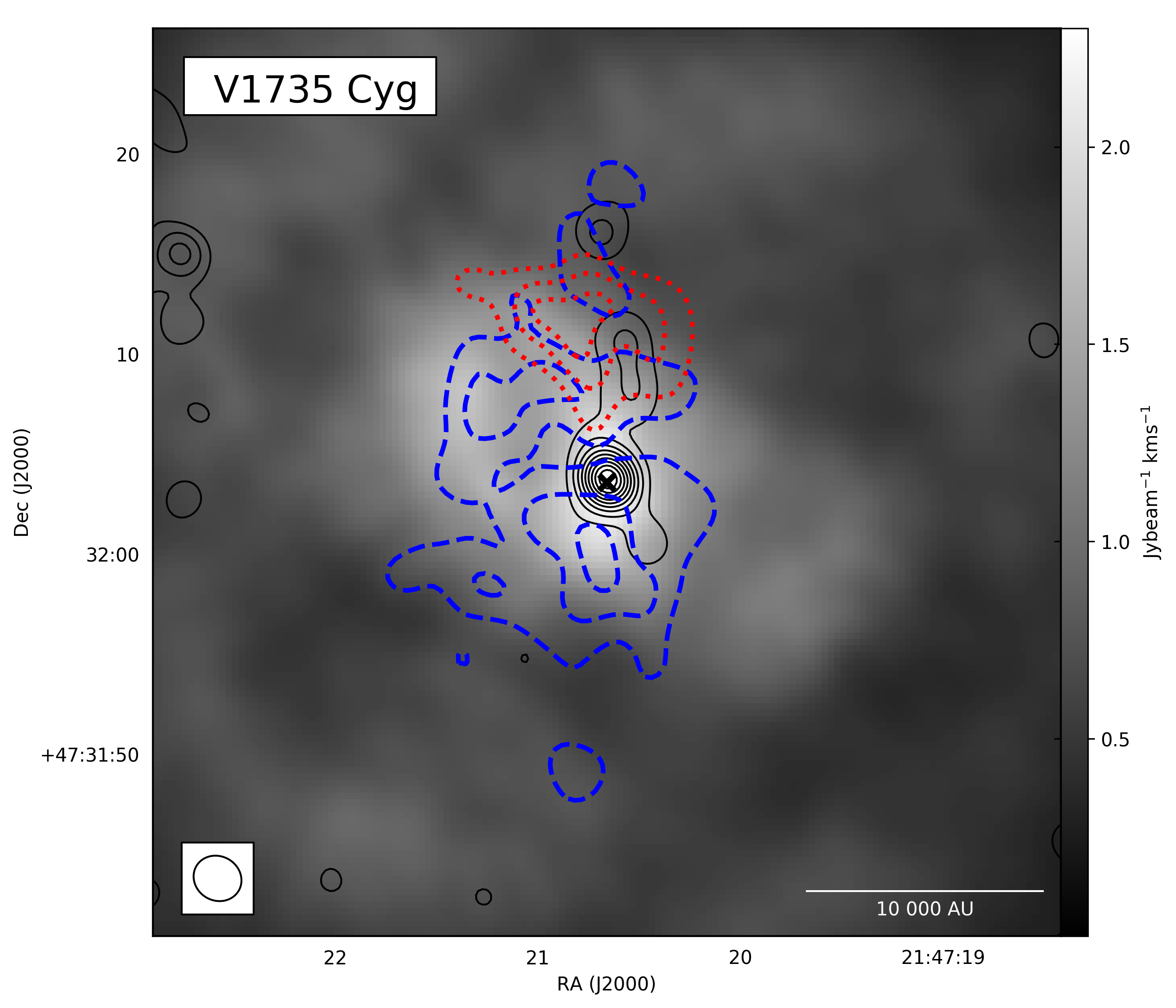}
  \includegraphics[width=.49\linewidth]{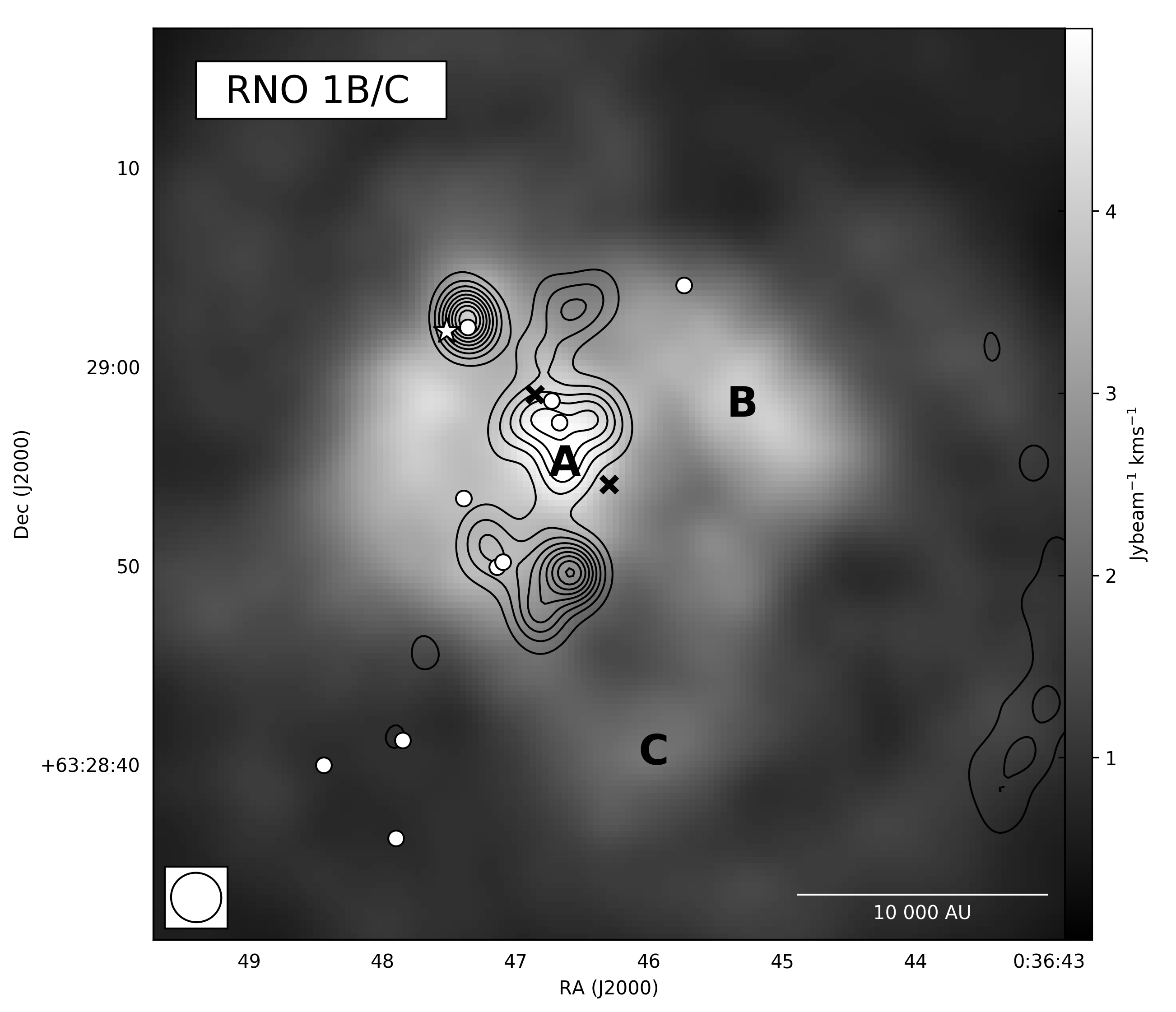}
 \caption{NOEMA CO and continuum maps of northern FUors. Left: $^{13}$CO integrated intensity map of V1735\,Cyg. The cross marks the optical position of the FUor, the black contours are the 2.7\,mm continuum emission in units of Jy\,beam$^{-1}$ at the 10...90\% of the peak. The dashed contour is the blue-shifted (integrated between -0.2--2\,km\,s$^{-1}$) and the dotted contour is the red-shifted lobe (integrated between 7--9.4\,km\,s$^{-1}$) of the outflow at 50, 70 and 90\% of their peak emission. Right: $^{13}$CO integrated intensity map of RNO\,1B/C. The black contours are the 2.7\,mm continuum emission at the at the 10...90\% of the peak. The two crosses are the members of the binary FUor-system (RNO\,1B in the center). The letters mark clumps identified on the CO map and circles mark other IR/submm/mm sources in the neighbourhood.}
   \label{fig1}
\end{center}
\end{figure}

The maps revealed extended, fragmented gas clouds around all our targets on a few thousand au spatial scale, with a local peak at or close to the star \cite[(Feh\'er et al. 2017)]{feher2017}. 
Only this central clump is found at e.g. V1057\,Cyg and V1735\,Cyg, but several other clumps were detected around e.g. V1515\,Cyg and V2492\,Cyg. The central clumps are roughly circular with radii of 1500--5000\,au and masses of 0.02--0.5\,M$_{\odot}$, often warmer than their surroundings, with average temperatures of 20--50\,K. Some of the CO clumps appear in the continuum and other continuum sources can also be seen close to some FUors. A good example is RNO\,1B/C (see Fig. \ref{fig1}, to the right). Clump\,A is located between the two members of the binary FUor and appears in both CO and continuum. Clump\,B is seen in CO but not seen in the continuum, however, the strong millimeter source to the south does not appear in CO. The other bright continuum source to the north-east appears as a faint local CO maximum. This source was also detected in the far-infrared. The velocity channel maps of V1057\,Cyg revealed a velocity gradient in the envelope, indicating rotation. A larger scale gradient  was measured in the environment of V2492\,Cyg, possibly caused by the shock front that appears as a bright H$\alpha$ rim close to the star. The $^{13}$CO lines around V1735\,Cyg have prominent line wings, indicating the presence of an outflow (see Fig. \ref{fig1}, to the left). The velocity channel maps around V1515\,Cyg show a feature similar to an expanding bubble around the star. Maps of V582\,Aur revealed that the star is located at the edge of an extended cloud structure with two well-identified velocity components and the unresolved continuum source around the FUor is possibly a disk with a total mass of 0.04\,M$_{\odot}$ \cite[(\'Abrah\'am et al. 2018)]{abraham2018}.

The classification of our targets into categories based on their evolutionary categories faces many issues. In the case of V1057\,Cyg the mass of the envelope corresponds to the characteristic mass of Class 0/I objects and it is detected in continuum, implying a significant spatial extent and dust content. However, the silicate feature appears in emission. This is only possible if there is no envelope present in the line of sight to absorb most of the 10\,$\mu$m emission or if the inclination of the system is high enough for the disk to be seen from the direction of an outflow cavity. The same is true for V1515\,Cyg, except its envelope is only weakly detected in the continuum. The lowest envelope mass was found at V2493\,Cyg and it was not detected in the 1.3\,mm or 2.7\,mm continuum either, suggesting a very small envelope and more evolved state. No 10\,$\mu$m spectra was measured for this FUor. For V733\,Cep and RNO\,1B/C there is no gas clump coinciding with the exact optical position of the star that can be identified as an envelope, but in both cases the silicate feature is detected in absorption. Neighbouring sources and their circumstellar structures might heavily influence the emission received from the targets.

\section{Future perspectives}

Recent studies by \cite[Cieza et al. (2018)]{cieza2018}, \cite[Li et al. (2017)]{li2017} and  \cite[Liu et al. (2017)]{liu2007} revealed that most FUors possess smaller disks than T\,Tauris. To sustain the observed high accretion rates these disks have to be really massive. Submillimeter Array data showed that some targets have very low spectral indices and their millimeter emission can be fitted with an inner, hot, optically thick disk and an outer, colder disk. Thus, a huge amount of mass may be hidden from the observer in the inner few astronomical units of these disks. Answering the question whether these small but massive FUor disks are evolutionary or intrinsic property will be addressed in the analysis of our ALMA Cycle\,4 observations.

The detailed analysis of the 2017 NOEMA observations for three targets is in progress and we expect to receive interferometric data of four other FUors (RNO\,127, PP\,13S, HH\,354 and V1184\,Tau) in the near future. This time, thanks to the new receiver PolyFiX not only CO but the emission of many other molecular species will be imaged on interferometric spatial resolutions. We will also propose observations for the more extended configuration of NOEMA for more close-by targets., since with an angular resolution of around 1$''$ we will be able to resolve the envelope-disk interface and its dynamics.

We are in the process of analysing about 40 molecular transitions of e.g. CN, SO, HC$_3$N, N$_2$H$^+$ in the 3\,mm band measured by the IRAM\,30m telescope. The data allows us to examine circumstellar morphology tracing structures of different density and chemical history. We aim to use the chemical model MUSCLE \cite[(Multi Stage CLoud codE, Semenov et al. 2010)]{semenov2010} to fit the observed density of the different molecular species and derive the physical parameters of the envelopes and other circumstellar structures. Using the models we can derive the time-dependent changes of molecular abundances and even infer the time and other characteristics of the FUor outburst. Since none of the FUor eruptions have been observed from their beginning to their end yet, this would give us important predictions for both new observations and the modeling of the FUor eruption process.

\section{Acknowledgement} 

This project has received funding from the European Research Council (ERC) under the European Union's Horizon 2020 research and innovation programme under grant agreement No 716155 (SACCRED, PI: \'A. K\'osp\'al).

\begin{discussion}
\discuss{Khaibrakhmanov} {Are there data on magnetic fields in the investigated objects?}
\discuss{Feher} {As far as we know there is no magnetic field measurements for our targets. For erupting sources we cannot measure the stellar photosphere but we have used Zeeman Doppler Imaging to measure the magnetic field of EX\,Lupi, which is a quiescent EXor.}
\end{discussion}

\end{document}